\documentclass[12pt]{article}
\usepackage{geometry}
\usepackage[utf8]{inputenc}
\usepackage{parskip}
\usepackage{verbatim}
\usepackage{amsmath}
\usepackage{amssymb}
\usepackage{graphicx}
\usepackage{caption}
\usepackage{subcaption}
\usepackage{hyperref}
\usepackage{booktabs}
\usepackage{bm}

\makeatletter

\newcommand{\lyxaddress}[1]{
	\par {\raggedright #1
	\vspace{1.4em}
	\noindent\par}
}
\usepackage{xcolor}

\makeatother

\begin{document}
\global\long\def\calH{\mathcal{H}}%
\global\long\def\calL{\mathcal{L}}%
\global\long\def\calM{\mathcal{M}}%
\global\long\def\calA{\mathcal{A}}%

\global\long\def\va{\bm{a}}%
\global\long\def\vb{\bm{b}}%
\global\long\def\vg{\bm{g}}%
\global\long\def\vj{\bm{j}}%
\global\long\def\vk{\bm{k}}%
\global\long\def\vn{\bm{n}}%
\global\long\def\vp{\bm{p}}%
\global\long\def\vq{\bm{q}}%
\global\long\def\vr{\bm{r}}%
\global\long\def\vs{\bm{s}}%
\global\long\def\vu{\bm{u}}%
\global\long\def\vv{\bm{v}}%
\global\long\def\vw{\bm{w}}%
\global\long\def\vx{\bm{x}}%
\global\long\def\vy{\bm{y}}%
\global\long\def\vz{\bm{z}}%

\global\long\def\vA{\bm{A}}%
\global\long\def\vB{\bm{B}}%
\global\long\def\vD{\bm{D}}%
\global\long\def\vE{\bm{E}}%
\global\long\def\vF{\bm{F}}%
\global\long\def\vH{\bm{H}}%
\global\long\def\vJ{\bm{J}}%
\global\long\def\vK{\bm{K}}%
\global\long\def\vL{\bm{L}}%
\global\long\def\vN{\bm{N}}%
\global\long\def\vP{\bm{P}}%
\global\long\def\vR{\bm{R}}%
\global\long\def\vS{\bm{S}}%
\global\long\def\vU{\bm{U}}%
\global\long\def\vX{\bm{X}}%

\global\long\def\val{\bm{\alpha}}%
\global\long\def\vom{\bm{\omega}}%
\global\long\def\vga{\bm{\gamma}}%
\global\long\def\vep{\bm{\epsilon}}%
\global\long\def\vnabla{\bm{\nabla}}%
\global\long\def\vmu{\bm{\mu}}%
\global\long\def\vnu{\bm{\nu}}%
\global\long\def\vsi{\bm{\sigma}}%
\global\long\def\vSi{\bm{\Sigma}}%

\global\long\def\order#1{\mathcal{O}\left(#1\right)}%
\global\long\def\edge#1{\left.#1\right|}%
\global\long\def\d{\mbox{d}}%
\global\long\def\bra#1{\left\langle #1\right|}%
\global\long\def\ket#1{\left| #1 \right\rangle }%
\global\long\def\G{\widetilde{G}}%
\global\long\def\tr{\mbox{Tr}}%
\global\long\def\Li{\mbox{Li}_{2}}%
\global\long\def\az{\alpha_{Z}}%
\global\long\def\ap{\alpha_{\pi}}%
\global\long\def\za{Z\alpha}%
\global\long\def\Ep{E_{\bm{p}}}%
\global\long\def\Eb#1{E_{{\scriptscriptstyle \text{bind},#1}}}%
\global\long\def\bs{\blacksquare}%
\global\long\def\GF{G_{{\scriptscriptstyle \text{F}}}}%

\title{Spin in Uniform Gravity, Hidden Momentum, and the Anomalous Hall Effect}
\author{Andrzej Czarnecki and Ting Gao}
\maketitle

\lyxaddress{Department of Physics, University of Alberta, Edmonton, AB,
Canada T6G 2E1}

\begin{abstract}
    We review the recent discussion of the absence of spin Hall effect in a uniform gravitational field, pointing out differences from the anomalous spin Hall effect in ferromagnetics despite a similar form of the Hamiltonian. 
\end{abstract}

\section{Introduction}\label{sec:intro}

Spin–orbit phenomena probe how internal angular momentum couples to background fields through the particle’s motion (its orbital degree of freedom). In condensed matter, this coupling produces, for example, the anomalous Hall effect (AHE): a transport current transverse to an applied electric field in ferromagnetics, described by an antisymmetric conductivity \(\sigma_{ij}^{\rm H}\), analysed by Karplus and Luttinger (KL) \cite{karplus1954hall} and subsequently interpreted in a Berry-curvature formulation (see \cite{yao2004first} and references therein; see also \cite{FangScience,Solovyev2003}; excellent reviews are \cite{NagaosaRMP2010,XiaoRMP2010,RevModPhys.87.1213}). In contrast, a recent claim of a “gravitational spin Hall effect” (SHE) in a uniform field for Dirac wave packets asserts a polarization-dependent transverse deflection without any transport geometry (no driven current) \cite{Wang:2023bmd}.

Our recent analysis  \cite{universe11110365} resolves this tension. First, an object with spin $\vS$ in a uniform gravitational field $\vg$ carries a hidden momentum \(\mathbf{p}_{\rm hidden}\sim \vS\times \vg/c^{2}\) that modifies the relation between canonical momentum and velocity. Second, for Dirac wave packets evolved with the Foldy–Wouthuysen (FW) Hamiltonian in a linear potential, the canonical transverse momentum is conserved while the velocity operator contains a spin term. The “at rest” initial state with \(\langle \mathbf{p}\rangle=0\) in  \cite{Wang:2023bmd} is therefore not at mechanical rest; preparing true rest requires a spin-dependent phase in the initial packet. With that preparation, any \(O(g)\) transverse drift vanishes. Third, using a FW transformation (linear in \(g\), all orders in \(1/c\)), we showed that spin-dependent transverse motion can appear no earlier than \(O(g^{2})\) for a broad class of states. These points together exclude a gravitational SHE at linear order in a uniform field.

A second theme of this note is to contrast this outcome with the KL AHE. The KL current is a linear-response, preparation-independent transport effect in crystals: \(J_i=\sum_j\sigma_{ij}E_j\), with an intrinsic contribution 
whose microscopic origin relies on Bloch periodicity: the interband matrix elements of the periodic part of \(-e\mathbf{E}\!\cdot\!\mathbf{r}\) (KL’s \(\calH''_{1}\)) are essential. In a uniform gravitational field there is no lattice periodicity (the natural basis is plane waves), so the KL objects \(J^{nn'}_b(\mathbf{k})\) and \(\calH''_{1}\) have no counterpart and the intrinsic Hall pathway is absent.

Below we summarise the hidden-momentum mechanism and its role in uniform gravity, restate and streamline the \(O(g)\) no-go result, and we contrast elements of the KL mechanism with the uniform gravity situation,  highlighting the missing ingredients for Hall transport in the latter.

\section{What constitutes a Hall effect?}\label{sec:Hall-ingredients}

A Hall effect is a transverse transport response to a longitudinal drive, involving, among other characteristics,
\begin{enumerate}
\item Driver and linear response. A uniform field \(\mathbf{E}\) drives a current \(J_i=\sum_j \sigma_{ij}E_j\); the antisymmetric part \(\sigma^{\rm H}_{ij}=\frac12(\sigma_{ij}-\sigma_{ji})\) is responsible for the  Hall effect.
\item Transport coefficient. The transverse deflection is caused by \(\sigma^{\rm H}_{ij}\) (or \(\rho_{ij}\)), not by an initial-state-dependent initial velocity.
\end{enumerate}
These conditions are satisfied by KL’s AHE but fail in a uniform gravitational field for neutral Dirac packets, where neither a transverse driver nor \(\sigma_{xy}\) are present.

\section{Spinning particle in gravitational field}
\label{sec: uniform g}

In this section, we use a simple model to show that a spinning object gains a ``hidden momentum" that modifies the relation between velocity and momentum in a gravitational field. However, 
the trajectory of the object in a uniform gravitational field is not modified by the spin.

\subsection{The hidden momentum}
\begin{figure}[h]
\centering\includegraphics[scale=0.6]{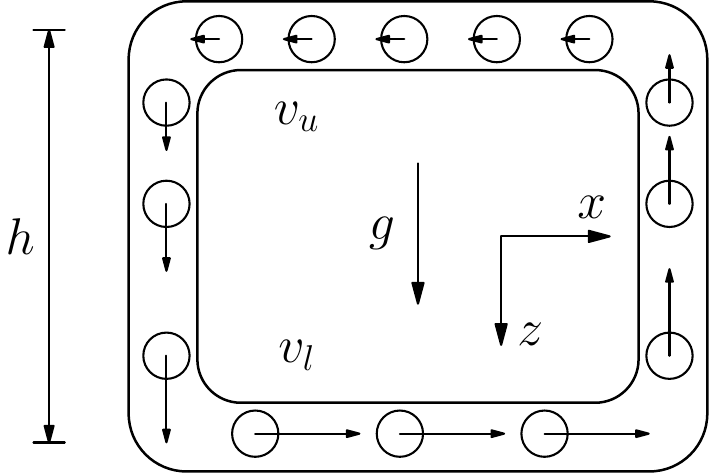}

\caption{Masses circulating in a frictionless pipe bent into a rectangle. The
rectangle is placed in the constant gravitational field with field strength $g$ near the Earth.\label{fig:Circul}}
\end{figure}

The idea of the hidden momentum first appeared in a 1967 paper
by Shockley and James \cite{shockley1967try} in the context of electromagnetism.
A concise discussion and a list of key references up to 2012 can be
found in~\cite{griffiths2012resource}. Following the electromagnetism discussion in~\cite{babson2009hidden}, we consider a circulating mass model in a gravitational field, as shown in Fig.~\ref{fig:Circul}.

In the rectangular pipe, the upper section contains $n_u$ masses moving with speed $v_u$ and momenta $p_u$, the lower section contains $n_l$ masses moving with speed $v_l$ and momenta $p_l$, and the masses accelerate/de-accelerate in the left/right vertical sections which connect the upper and lower pipes. Current conservation imposes a condition on the velocity $n_uv_u=n_lv_l$, whereas the total momentum reduces to a difference between the momenta in the lower and upper section $p_{\mathrm{tot}}=n_lp_l-n_up_u$. Without relativistic effects, the relation $p=mv$ guarantees that the momentum within the pipe adds up exactly to zero; however, when relativity is accounted for, the momenta gain an extra factor $\gamma$ due to gravitational and special relativity time dilation:
\begin{align}
    p_u&=m\gamma_u v_u,\quad p_l=m\gamma_l v_l,\\
    \gamma &=\frac{1}{\sqrt{\left(1-\frac{gz}{c^{2}}\right)^{2}-v^{2}/c^{2}}}\\
    &\simeq1+\frac{gz}{c^{2}}+\frac{v^{2}}{2c^{2}},
\end{align}
where the origin is chosen at the center of the rectangular loop, and the $z$-direction is along the same direction as $\boldsymbol{g}$. With the difference in $\gamma$ between the upper and lower section, the total momentum gains a ``hidden'' $x$-component, in the sense that does not vanish even when there is no overall linear motion of the system:
\begin{align}
    p_{\mathrm{hidden}}&=n_lmv_l\left( \frac{gh}{c^2}+\frac{v_l^2-v_u^2}{2c^2}\right)\\
    &=2n_lmv_l\frac{gh}{c^2}.
\end{align}
To manifest the connection between angular motion and hidden momentum, the result can be rewritten in terms of the total angular momentum of the system:
\begin{equation}
    \boldsymbol{p}_{\mathrm{hidden}}=\frac{\boldsymbol{L}\times\boldsymbol{g}}{c^{2}}.\label{eq:pmech}
\end{equation}
\subsection{Trajectory in a uniform gravitational field}
It is claimed in~\cite{Wang:2023bmd} that particles carrying spin will be deflected in a spin-dependent way in a uniform gravitational field. Here we take another look at the discussion from the viewpoint of hidden momentum.

In~\cite{Wang:2023bmd}, the Dirac Hamiltonian coupled to the uniform gravitational field is Foldy-Wouthuysen-transformed into a non-relativistic single particle Hamiltonian at leading order in $g$. For positive energy states, this Hamiltonian is written as
\begin{align}
    \mathcal{H}_{\mathrm{FW}}=& Vmc^2+\frac{\boldsymbol{p}\cdot \left(V\boldsymbol{p}\right)}{2m}\\
    &+\frac{g\hbar}{4mc^2}\left(\sigma_xp_y-\sigma_yp_x\right),\\
    V=& 1-\frac{gz}{c^2}.
    \label{Hamiltonian, LO}
\end{align}
By evaluating the time-evolution of the expectation value of the position operators for a Gaussian wave packet with $\langle\boldsymbol{x}(0)\rangle=0$, $\langle\boldsymbol{p}(0)\rangle=0$, and $\langle\sigma_y\rangle=\pm1$, \cite{Wang:2023bmd} concluded that particles with spin along the $\pm y$ direction are deflected in the $x$ direction with the velocity $v_x=\mp g\hbar/(4mc^2)$. However, as shown in~\cite{universe11110365}  a look at the classical Hamilton's equations
\begin{align}
    m\frac{dx}{dt}&=m\frac{\partial \mathcal{H}_{\mathrm{FW}}}{\partial p_x}\\
    &=Vp_x-\frac{g\hbar}{4c^2}\sigma_y.
\label{eq:23}
\end{align}
shows that the particle carries a hidden momentum, which indicates that the deflection is due to a nonzero initial velocity rather than a dynamical evolution of the wave packet.

\begin{figure}[htb]
 \centering
     \includegraphics[scale=.4]{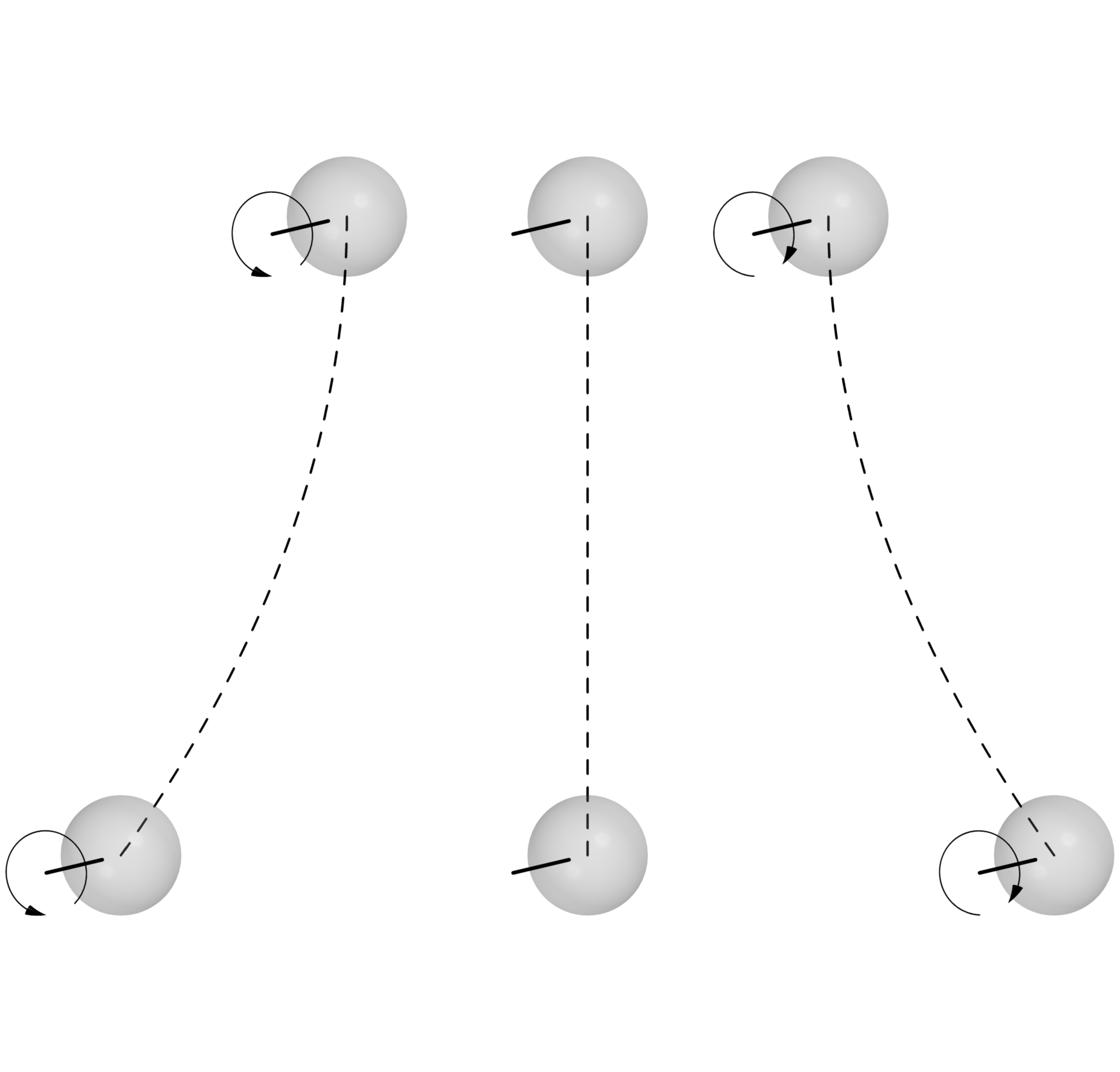}\\[-4mm]
     (a)\\
  \includegraphics[scale=.4]{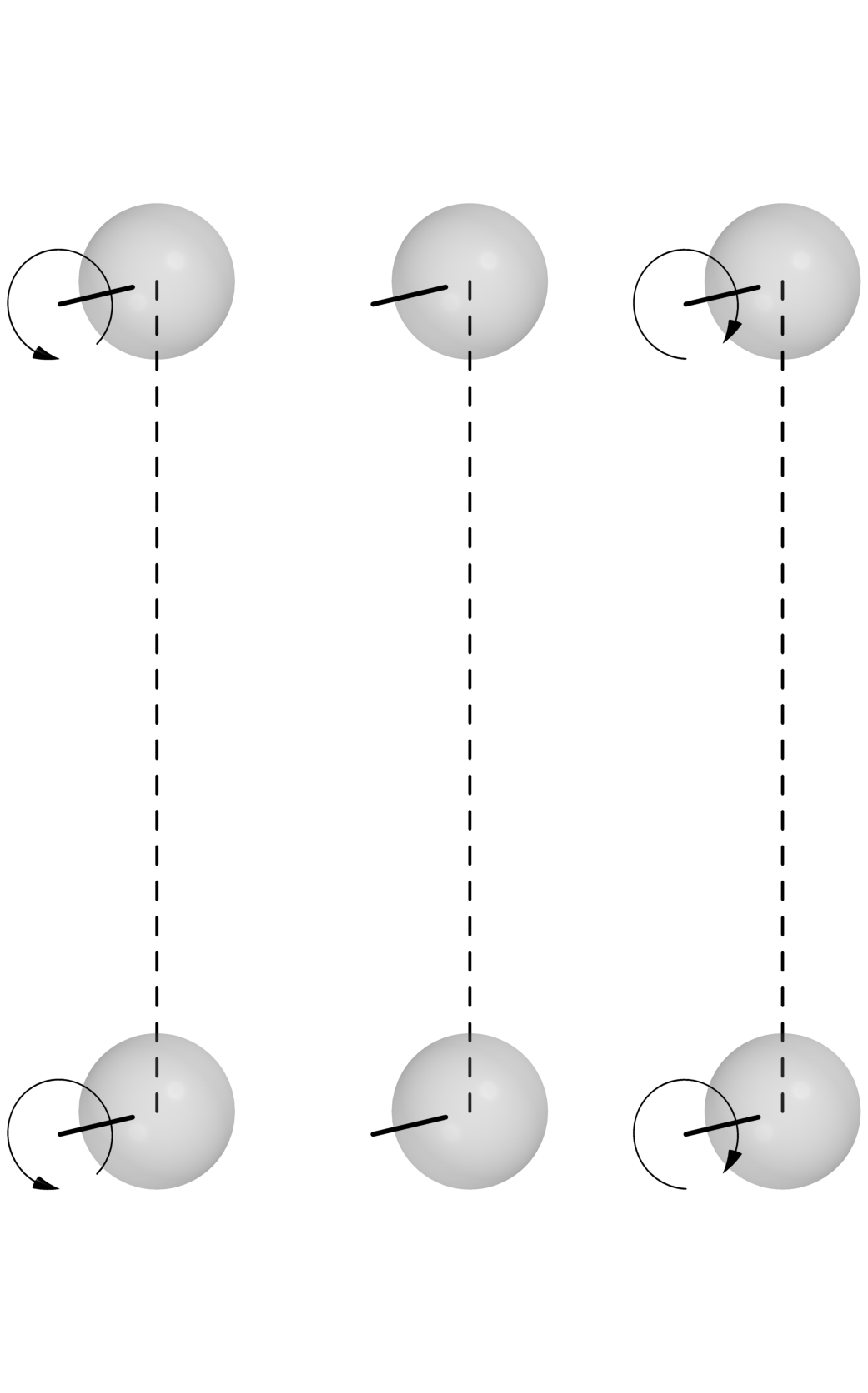}\\[-4mm]
(b)
    \caption{Trajectories for particles with spin $\sigma_y=+1$ (left), $\boldsymbol{\sigma}=0$ (middle), and $\sigma_y=-1$ (right) when (a) the trajectories experience a spin-dependent deflection; and (b) the trajectories are independent of the spin. The initial velocity is along the $z$-direction (vertical) and is chosen to be the same for all  particles for the ease of comparison.}
    \label{fig:trajectory}
\end{figure}

Further, we show that this spin-dependent deflection of the trajectory is unobservable in a free fall experiment, to the order in $g$ to which the Hamiltonian~\eqref{Hamiltonian, LO} is valid. To illustrate our point, we consider the trajectory of three particles: two Dirac particles with the spin along the $\pm y$ direction, and one scalar which does not carry spin. With the three trajectories initially parallel to each other and carrying the same initial velocity along the $z$ direction, a spin dependent deflection would cause the trajectories to deflect from each other, as shown in Fig.~\ref{fig:trajectory}(a); however, the condition the three trajectories are initially parallel to each other implies
\begin{equation}
    p_x(0)=\frac{g\hbar}{4c^2}\sigma_y
\end{equation}
for the Dirac particles, and
\begin{equation}
    p_x(0)=0
\end{equation}
for the scalar particle. Therefore, due to the cancellation between the initial (total) momentum and the hidden momentum, the subsequent motions of the three particles are completely identical, as shown in Fig.~\ref{fig:trajectory}(b). While we discuss the classical trajectory for simplicity, a complete analog of the discussion can be obtained in quantum mechanics as in \cite{universe11110365}.

\section{Comparison with Hall effects in Ferromagnetics}
While the spin-orbit coupling in the uniform gravitational field does not change the orbit, there has been a belief since the 1950s that a similar type of coupling is responsible for the anomalous Hall effect in ferromagnetics~\cite{karplus1954hall}.
In this section we briefly review this mechanism and discuss the difference in the consequences.

As considered in \cite{karplus1954hall}, the Hamiltonian for an electron in ferromagnetics with spin polarized along the direction of the magnetization $\boldsymbol{M}$ in an external electric field $\boldsymbol{E}$ can be written as
\begin{align}
    \mathcal{H}_T &=\mathcal{H}_0+\mathcal{H}'+\mathcal{H}'',\\
    \mathcal{H}_0 &=\frac{p^2}{2m}+U(\boldsymbol{r}),\\
    \mathcal{H}' &=\frac{1}{4m^2c^2}\frac{\left(\boldsymbol{M}\times\boldsymbol{\nabla} U\right)\cdot\boldsymbol{p}}{M_s},\\
    \mathcal{H}''&=-e\boldsymbol{E}\cdot\boldsymbol{r},
\end{align}
where $U(\boldsymbol{r})$ is the periodic crystal potential, $M_s$ is the maximum magnetization,
and $e$ is the charge of the electron. Under the eigenbasis $\phi_{(n,\boldsymbol{k})}$ of $\mathcal{H}=\mathcal{H}_0+\mathcal{H}'$, $\mathcal{H}''$ is found to have both a singular part $\mathcal{H}_i''$ responsible for the usual conductivity effects, whose matrix element is 
\begin{equation}
    \langle n\boldsymbol{k}\vert \mathcal{H}_1''\vert n' \boldsymbol{k}'\rangle=-ie\delta_{nn'}\boldsymbol{E}\cdot\boldsymbol{\nabla}_{k}\delta_{\boldsymbol{kk}'},
\end{equation}
and a regular periodic part $\mathcal{H}_1''$ responsible for the anomalous Hall effect which gives the matrix element
\begin{equation}
    \langle n\boldsymbol{k}\vert \mathcal{H}_1''\vert n' \boldsymbol{k}'\rangle=-ie\delta_{\boldsymbol{kk}'}\boldsymbol{E}\cdot\boldsymbol{J}^{nn'},
\end{equation}
where
\begin{equation}
    \boldsymbol{J}^{nn'}(\boldsymbol{k})=\int w_{n\boldsymbol{k}}^*(r)\boldsymbol{\nabla}_{k}w_{n'\boldsymbol{k}}(r)d^3x,
\end{equation}
and $\phi_{(n,k)}=e^{i\boldsymbol{k}\cdot\boldsymbol{r}}w_{n\boldsymbol{k}}(\boldsymbol{r})$. Averaging the velocity over the Fermi distribution $\rho$ then gives
\begin{equation}
    \bar{v}_a=-ieE_b\sum_l \rho_0'(\epsilon_l)v_b(l)J_a(l),
\end{equation}
where $a$ and $b$ are vector indices, $\rho_0$ is the distribution with respect to $\mathcal{H}$,  $\rho_0'=\partial\rho_0/\partial\epsilon$, and $J_a(l)=J_b^{nn}(k)$. The average velocity here then gives rise to a transverse current $\boldsymbol{J}=N_de\bar{\boldsymbol{v}}=r\boldsymbol{M}\times\boldsymbol{E}$, where the coefficient $r$ depend on the material peroperties.

Comparing with the discussion for the spin-orbit coupling in uniform gravitational field in Sec.~\ref{sec: uniform g}, we note that, different from ferromagnetics where the lattice potential $U(\boldsymbol{r})$ gives rise to a non-trivial eigenbasis $\phi_{(n,\boldsymbol{k})}$, the eigenbasis for the Hamiltonian in Eq.~\eqref{Hamiltonian, LO} in the absence of the gravitational field are plane waves. Therefore, the regular periodic perturbation $\mathcal{H}_1''$ does not have a counter part in the uniform gravitational field. Indeed, as already noted in~\cite{karplus1954hall}, when the band becomes almost free, its contribution to the average velocity becomes almost zero. Therefore, while these two effects look very similar, the transverse velocity in ferromagnetics does not occur in the uniform gravitational field due to the absence of the essential lattice potential.

\section{Summary}
In this note, we reviewed the recent discussion for the hidden momentum and the trajectory of Dirac particles in the uniform gravitational field, and compared the results obtained therein with the Hall effects in ferromagnetics.

Using a circulating mass model, we have shown that a spinning object obtains a hidden momentum in a gravitational field and this hidden momentum modifies the relation between velocity and canonical momentum. However, with a given trajectory, it is not possible to tell the spin of the particle, therefore, up to the linear in $g$ approximation adopted here, the spin Hall effect does not exist in the uniform gravitational field in the usual sense. While a similar Hamiltonian leads to the anomalous spin Hall effect in ferromagnetics, the lattice potential is crucial for this phenomenon and the same conclusion does not apply to free particles in the uniform gravitational field.

\subsection*{Acknowledgments}

We used Chat GPT 5 for literature research and improving text. We used
Asymptote \cite{Asymptote} to make drawings. This research was funded
by Natural Sciences and Engineering Research Canada (NSERC).

\end{document}